# Electron-beam-controlled volatile nanomechanical bistability

Toji Thomas, Kevin F. MacDonald and Eric Plum

Optoelectronics Research Centre, University of Southampton, Highfield, Southampton, SO17 1BJ, UK

**Abstract:** Bistability in nanomechanical resonators can be exploited for sensing, signal processing, and memory applications due to its potential for switching and high sensitivity to external stimuli. External vibration can be used to drive a doubly-clamped nanowire into the nonlinear regime of bistable oscillation. Here, we experimentally demonstrate that the bistable oscillation of such a nonlinear nanomechanical resonator can be controlled, switched and read by an electron beam.

## Introduction

Systems that can exhibit two distinct output states in response to the same input, depending on the history of previous inputs, are known as bistable. Bistability is a feature of nonlinear systems and many systems become nonlinear upon sufficiently strong excitation. Consider a typical harmonic oscillator, a doubly-clamped nanowire. Increasing lateral displacement will stretch the nanowire increasingly rapidly, causing a stiffening effect that gives rise to a cubic nonlinearity and bistability. The bistable states correspond to different amplitudes of oscillation, and the input may be an external vibration driving its oscillation or any other physical parameter that the vibration depends on. Bistable states can serve as memory and a high-contrast transition between nanomechanical bistable states in response to small changes provides opportunities for highly sensitive detection, e.g. of forces, charges and masses [1-3]. Nanomechanical bistable states can be driven or affected by external and internal factors such as external force[4], capacitance[5], magnetic field[6], stochastic noise[7] or optomechanical coupling[8]. In analogy to optical bistability, which enables ultrafast all-optical switching [9], memory elements[10], pulse shaping and regeneration[11], logic gates[8], transistors and amplifiers[12] etc, all-mechanical bistable signal processing functionalities may also be envisaged, e.g. in the context of the emerging areas of time crystals[13] and timetronics[14]. Exploiting that optical heating can shift nanomechanical resonances[15], the concepts of optical and mechanical bistabilities have been combined to realize a mechanical bistability controlled by light, using nanowires decorated with plasmonic metamolecules[16].

Here, we report that the bistable mechanical states of a doubly-clamped nanowire, a fundamental and nonlinear building block of NEMS, MEMS and photonic metamaterials[13, 17, 18], may be controlled by an electron beam. As introduced above, when driven into the resonant and nonlinear regime of mechanical oscillation, a nanowire can support two different amplitudes of oscillation under identical conditions. Excitation of and switching between the bistable states may be controlled by changing the frequency of external vibration relative to the nanowire's fixed resonance frequency, or by changing the nanowire's resonance frequency relative to a fixed frequency of external vibration. We show that the latter can be achieved with an electron beam as follows. Energy transfer from electron impact on the nanowire raises its temperature, and the associated thermal expansion reduces the nanowire's stress, decreasing its mechanical resonance frequencies. For a nanowire that is mechanically driven into the bistable regime of oscillation, the electron-beam-induced resonance frequency changes can be used to cycle through the hysteresis loop of bistable oscillation, i.e. to switch between the bistable states of high and low amplitude oscillation, which are stable under identical conditions (Figure 1). This may be achieved by varying the proximity of the electron beam from the edge of the nanowire, or by modulating the electron beam current. As the electron beam can be used to control the nanowire's resonance frequency, it also provides control over the frequency of mechanical oscillation at which bistability occurs.

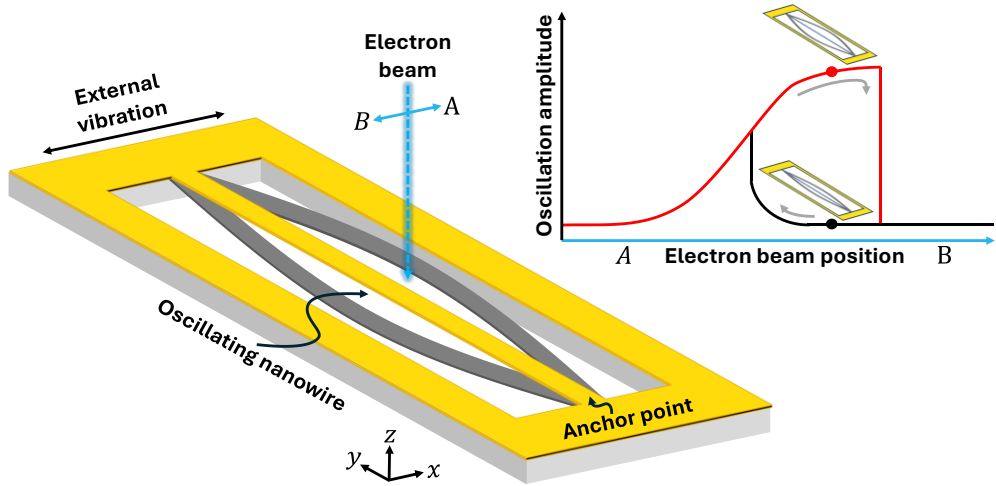

*Figure 1: A free-standing nanowire is driven to its bistable regime of in-plane mechanical oscillation. The nanowire oscillation and switching between its bistable states of oscillation may be controlled by moving an electron beam relative to the nanowire, as illustrated by the hysteresis loop. Based on the history of the electron beam position the nanowire attains a high or low amplitude of stable mechanical oscillation under identical conditions.*

**Observation of bistability using an electron beam**

The nanowire used here has 47 μm length (L), 400 nm width (W) and a thickness (H) of 150 nm (Figure 2a). It was fabricated by focused ion beam milling (FEI Helios Nanolab 600) of a 50 nm thick silicon nitride membrane supported by a 200 μm thick silicon frame (NORCADA) that was coated with a 50 nm of thermally evaporated (BOC Edwards) gold and then 50 nm of sputtered (AJA Orion) gallium nitride (GaN). The nanowire is anchored to the silicon frame at both ends. The entire sample is mounted on a shear piezoelectric stack (Thorlabs PL5FBP3) to drive in-plane displacements along $x$ (Figure 2a). Piezo displacements are driven by a sinusoidal voltage with swept (or fixed) frequency from the signal output of a lock-in amplifier (Zurich Instruments UHFLI 600 MHz). The experiment was conducted using an SEM (CamScan3600). The sample is kept inside the SEM chamber in vacuum with 3 μbar pressure. Lock-in detection of the modulation of the secondary electron signal arising from nanowire motion relative to the electron beam position was used to study the nanowire's mechanical response[19-21]. The SEM is operated in 'spot mode,' where the electron beam with 1.3 nA current ($I_e$) and 10 kV acceleration voltage ($V_e$) is positioned (blue dot in Figure 2a) near the centre of the nanowire (along $y$) at a distance (d) away from its edge (along $x$). The secondary electron signal is collected using an Everhart-Thornley secondary electron detector and fed into the signal input channel of the lock-in amplifier for detection, using the driving signal as the reference.

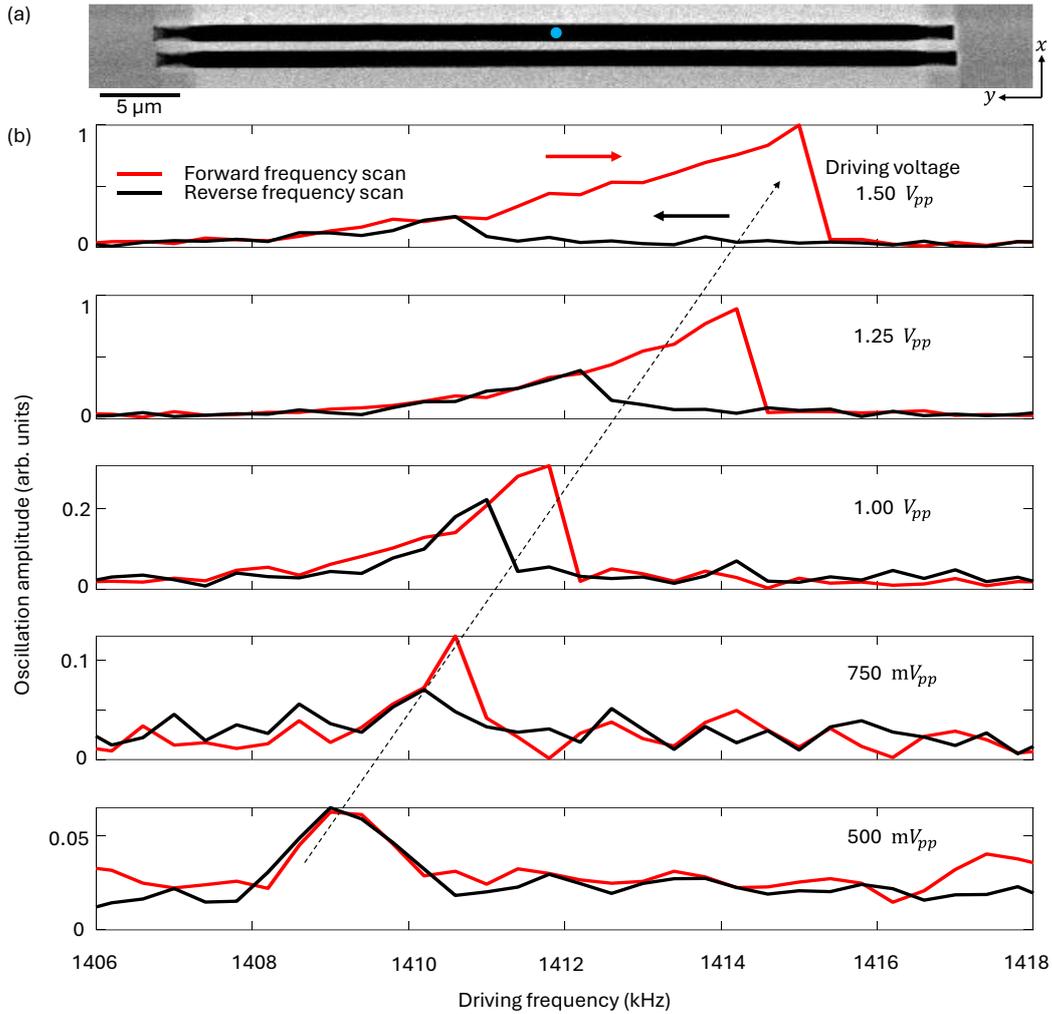

*Figure 2: The nanowire and its transition from linear to non-linear oscillation. (a) SEM image of the nanowire. In-plane displacement along x is driven by a piezoelectric transducer supplied with a sinusoidal driving voltage and detected as modulation of the secondary electron signal with a fixed electron beam impact position (blue dot) about d = 150 nm from the nanowire edge. (b) Amplitude of nanowire oscillation as a function of increasing (red) and decreasing (black) driving frequency for different driving voltage amplitudes. The nanowire transitions from linear to non-linear and bistable oscillation with increasing driving amplitude.*

Figure 2b shows how the nanowire transitions from linear to nonlinear and bistable oscillation when being shaken with increasing amplitude. The oscillations are driven via the piezo using bidirectional frequency sweeps at different fixed voltage amplitudes from 500 m$V_{pp}$ to 1.5 $V_{pp}$. The electron beam passes at a distance d = 150 nm from the nanowire edge and the oscillation of the nanowire's centre is monitored by detecting the resulting modulation of the secondary electron signal. To show the nanowire's amplitude of oscillation relative to its driven anchor points, the anchor point oscillations given by the average non-resonant measurement background were subtracted. The curves are normalized to the overall maximum oscillation amplitude occurring for 1.5 $V_{pp}$. The bidirectional frequency sweeps at a low driving voltage (500 m$V_{pp}$) reveal Lorentzian bell shapes at the same frequency. This characteristic resonance peak of a harmonic oscillator evolves into an asymmetric resonance peak with a sharp high frequency foldover wing[22] and hysteresis as the amplitude of flexural oscillations increases at larger driving voltages. This leads to bistability: The shape of the curve depends on the frequency sweep direction, and the non-linear resonator can have a large or small amplitude of oscillation under identical conditions depending on the history previous driving frequencies. For example, at a driving voltage amplitude of 1.5 $V_{pp}$, the oscillation amplitude is 18 times larger when the oscillation frequency is increased to 1414 kHz than when it is decreased to 1414 kHz.

The bistability occurs due to the nanowire's mechanical nonlinearity, also known as geometric nonlinearity, which arises from the nanowire stiffening due to being stretched by significant lateral displacement as its anchor points remain fixed. The resulting motion can be described by the nonlinear Duffing equation[23], $\ddot{x} + \Gamma\dot{x} + \omega_o^2 x + \alpha x^3 = \frac{A_o}{m_{\text{eff}}}\cos(\omega t)$. Here, $\Gamma, \omega_o, m_{\text{eff}}$ are the damping parameter, in-plane resonance frequency and effective mass of the nanowire. $A_o$ is the amplitude of the external driving force that is oscillating at driving frequency $\omega$. Stiffening of the nanowire due to stretching increases its spring constant at larger displacements, which is accounted for by the cubic term, where $\alpha$ is the nonlinear stiffness coefficient. This Duffing nonlinearity becomes significant as the in-plane displacement $x$ approaches the nanowire's width.

**Effect of the electron beam on the bistable resonance**

Energy dissipated as the electron beam interacts with the nanowire by inelastic scattering of electrons increases the nanowire's temperature, leading to thermal expansion and associated changes in stress and resonance frequency. To investigate how the electron beam position relative to the nanowire affects the bistable response, we characterized the nanowire oscillation in response to increasing and decreasing driving frequencies at a fixed driving voltage of 1.5 $V_{pp}$ for different distances *d* between electron beam and nanowire edge (Figure 3). The oscillation amplitude signals are normalized to the maximum obtained at each electron beam location. The data show that the nanowire's resonance and associated bistable response red-shift as the electron beam is positioned closer to the nanowire edge. Such a shift may be expected to arise from electron-beam-induced heating of the nanowire, which will cause thermal expansion and an associated stress reduction, which will reduce the nanowire's natural frequency. The closer the electron beam is to the oscillating nanowire the more electrons will interact with the nanowire and the bigger temperature increase, stress reduction and red-shift. Qualitatively similar behaviour has been reported for optical heating of mechanical resonators[16]. For electron-beam-induced heating, the effect depends on the geometry, electron beam parameters and effective thermal conductivity, $\kappa$ of the nanowire. The average temperature change above ambient temperature of the nanowire can be estimated as $\Delta T \approx \xi P' \text{L}/(8\kappa \text{WH})$[23], where $P'$ is the amount of electron beam power falling on the nanowire and $\xi$ being the fraction of that power absorbed by the nanowire materials that will contribute to the heat production. To provide an upper limit, we shall assume that all incident power ($P' = I_e V_e$) is converted to heat ($\xi = 1$). With $\kappa \approx 156$ Wm$^{-1}$K$^{-1}$, the electron beam carries enough power to heat the nanowire by $\Delta T \leq 8$ K. Following [23], the cooling time is $\Theta = \text{L}^2 C_v/8\kappa \approx 4.5$ μs where $C_v$ =2.54 x $10^6$ $Jm^{-3}K^{-1}$ is the effective volume heat capacity of the nanowire.

This can be compared with an estimate of the temperature increase required to explain the observed frequency shift based on bulk material parameters. The nanowire's resonance frequency according to Euler–Bernoulli beam[24, 25] theory is given by $f_o(\sigma) = 1.03\left(\frac{\text{W}}{\text{L}^2}\right)\sqrt{\frac{E}{\rho}}\sqrt{1 + \frac{\sigma \text{L}^2}{3.4EW^2}}$, with effective density $\rho \approx 9490$ kg m$^{-3}$, effective Young's modulus $E \approx 205$ GPa, tensile stress $\sigma = \sigma_0 - \alpha_{\text{eff}} E \Delta T$, initial stress $\sigma_0$ and effective thermal expansion coefficient $\alpha_{\text{eff}} \approx 4.1$ x $10^{-6}$ K$^{-1}$ of the nanowire. Considering a resonance frequency $f_o = 1416.2$ kHz at *d* = 200 nm (Figure 3), where the nanowire is least thermally affected by the electron beam, we estimate the initial stress $\sigma_0$ to be 84.3 MPa. The resonance shift of -4.4 kHz to 1411.8 kHz that is observed as the electron beam is moved much closer to the nanowire edge (*d* = 50 nm) implies a stress reduction by about 0.8 MPa and an electron-beam-induced increase of the nanowire's temperature by $\Delta T \approx 1$ K, which is consistent with the heating the electron beam can provide.

Being dependent on the power carried by the electron beam, it can be expected that the temperature increase $\Delta T$ can also be controlled by the electron beam current and acceleration voltage. Indeed, a similar red-shift of the bistable resonance can be observed when increasing the beam current (Supplementary Figure S1).

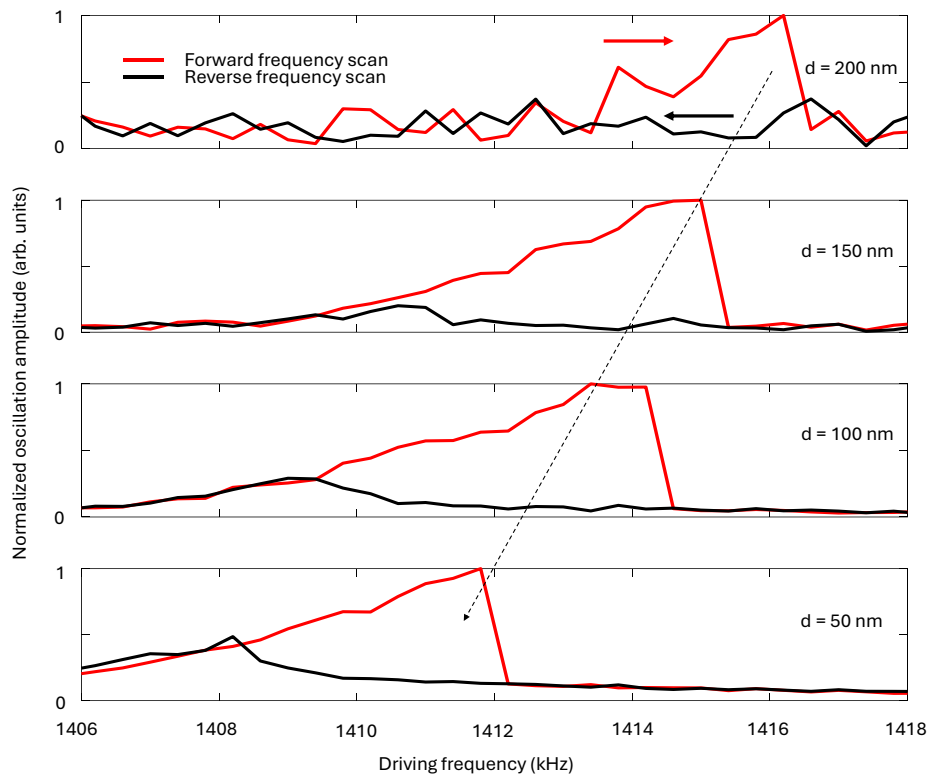

*Figure 3: Effect of the electron beam position on the nanowire's bistable resonance. Nanowire oscillation amplitude as a function of increasing and decreasing driving frequency for a fixed driving voltage (1.5 $V_{pp}$) and different distances* d *of the electron beam from the nanowire edge. The data for each distance is normalized to its own maximum. The black dashed arrow indicates a red-shift of the bistable resonance as the electron beam is positioned closer to the edge of the oscillating nanowire.*

**Electron-beam-controlled switching between bistable states**

Given that the spectral position of the bistable resonance depends on the position of the electron beam with respect to the nanowire, one may expect that switching between the bistable states can be achieved by modulating the electron beam position instead of the driving frequency. Figure 4 shows the nanowire's oscillation amplitude as a function of decreasing and increasing distance d between electron beam and nanowire for selected fixed driving frequencies. We performed bidirectional line scans of the electron beam across the nanowire edge, identifying the edge (d = 0) as the maximum gradient of the DC secondary electron signal (which is equivalent to finding the edge on an SEM image). Since the secondary electron signal modulation *M* at the driving frequency is proportional to both the nanowire's oscillation amplitude *A* and the local gradient *G* of the DC secondary electron signal in the direction of movement[19], the nanowire's oscillation amplitude was determined as the ratio *M/G*. We do indeed observe bistability of the nanowire's oscillation amplitude as the electron beam is moved relative to the nanowire edge, with the oscillation amplitude following the upper (lower) branch of the hysteresis loop when the electron beam moves towards (away from) the nanowire. At lower (higher) driving frequencies, the bistability is observed for electron beam positions that are closer to (further from) the nanowire.

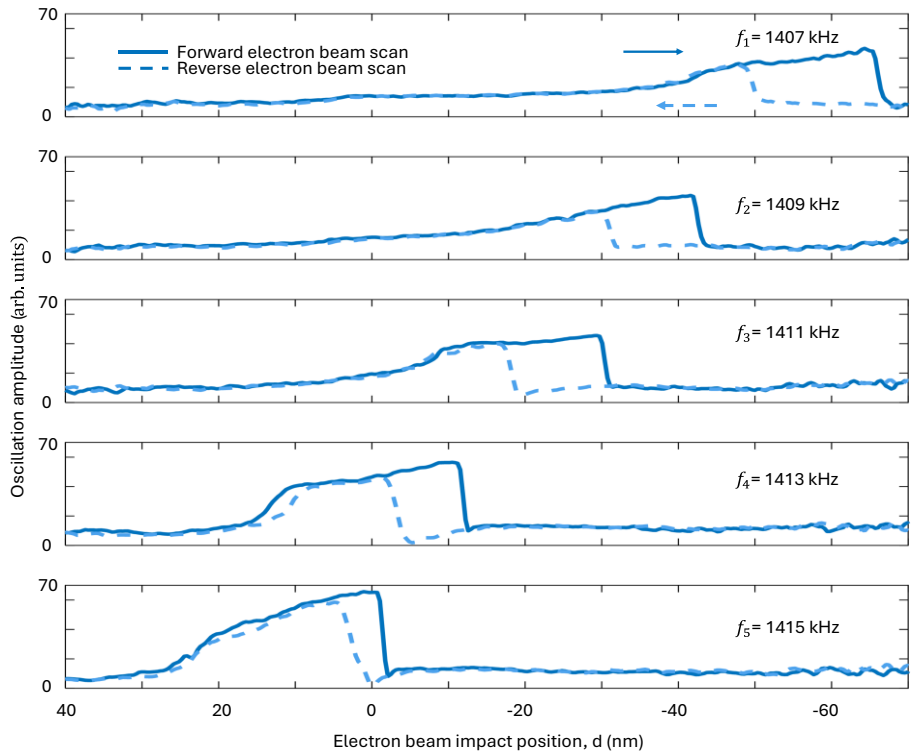

*Figure 4: Switching between bistable states using an electron beam. Nanowire oscillation amplitude as a function of decreasing (solid) and increasing (dashed) distance between electron beam and nanowire for selected driving frequencies. The nanowire edge is located at* d=0*. A hysteresis loop can be seen, where – for the same electron beam position – the nanowire can oscillate at very different amplitudes, depending on the history of previous electron beam positions.*

As illustrated by Fig. 3 and discussed above, the nanowire's mechanical resonance frequency depends on the distance between electron beam and nanowire, as any electrons impacting on the nanowire will raise its temperature, resulting in thermal expansion that lowers its tension and thus also its natural frequency. Therefore, as the electron beam approaches (moves away from) the nanowire, heating from more (cooling from fewer) electrons impacting on the nanowire shifts its mechanical resonance to lower (higher) frequencies. In Fig. 4, as the resonance is red-shifted (blue-shifted) past the fixed driving frequency, the oscillation amplitude follows the upper (lower) branch of the hysteresis loop.

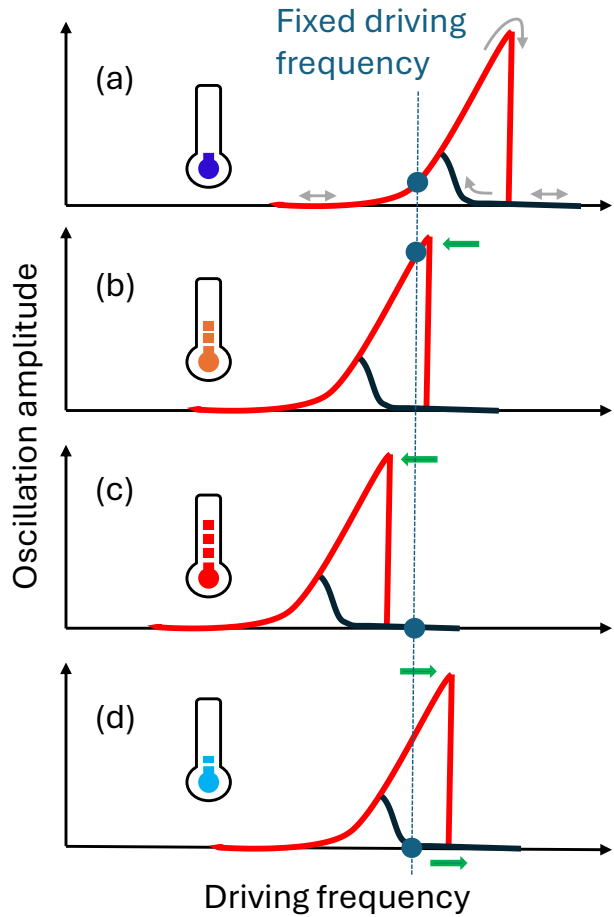

*Figure 5: Mechanism of electron beam control of nanomechanical bistability. The nanowire is driven externally by sinusoidal vibration to a highly nonlinear regime in which hysteresis of its oscillation amplitude can be observed at resonance. (a) The external driving frequency is set (blue dot) below the mechanical resonance while a distant electron beam provides little heating (thermometer). (b) By moving the electron beam closer to the nanowire, more electrons impact on and deposit energy in the nanowire, raising its temperature. Thermal expansion reduces the stress in the nanowire and the resonance shifts (green arrow) towards lower frequencies. The nanowire's oscillation amplitude follows the upper branch of the hysteresis loop. (c) With further electron-beam-induced temperature increase, the hysteresis edge shifts beyond the fixed driving frequency: the nanowire switches to low amplitude oscillation. (d) As the electron beam moves away from the resonator, less electron beam power is deposited in the nanowire, allowing it to cool and the resonance shifts towards higher frequencies. The nanowire oscillation amplitude follows the lower branch of the hysteresis loop, returning to situation (a).*

Figure 5 illustrates this step-by-step, starting with the electron beam far from the nanowire that is resonant above the fixed driving frequency (low temperature, Fig. 5a). As the electron beam approaches the nanowire (rising temperature), the resonance red-shifts leading to resonant driving of the nanowire with the oscillation amplitude following the upper branch of the hysteresis loop that is characterized by large nanowire oscillations (Fig. 5b). As the electron beam approaches the nanowire even more (highest temperatures) the resonance red-shifts past the driving frequency and the oscillation amplitudes drops to its low non-resonant level (Fig. 5c). When the electron beam moves away from the nanowire (falling temperature), the resonance blue-shifts and the nanowire's oscillation amplitude follows the lower branch of the hysteresis loop (Fig. 5d), until it the resonance frequency is above the fixed driving frequency (Fig. 5a).

This model also explains how the fixed driving frequency controls the electron beam position at which the bistability can be observed. The bistable states can be observed when the driving frequency matches the nanonwire's resonance frequency. The lower the driving frequency, the more electron-beam-induced red-shift of the nanowire's natural frequency is required to achieve this (i.e. to go from Fig. 5a to Fig. 5b), and a bigger red-shift occurs when the electron beam is positioned closer to the nanowire. Therefore, at the lowest driving frequency the bistability is observed when the electron beam is closest to the nanowire or indeed positioned on the nanowire (Fig. 4). As electron-beam-induced heating will be most sensitive to the electron beam position around the nanowire's edge, it may be expected that the nanowire's temperature and resonance frequency will most strongly depend on the electron beam position around $d=0$. It follows that the hysteresis loop as a function of electron beam position should be particularly narrow near the nanowire edge and indeed, this is what we observe when driving the nanowire at 1415 kHz.

**Conclusion**

In conclusion, we have demonstrated how the bistable oscillation of a nonlinear nanomechanical resonator – a nanowire – can be controlled, switched and read by an electron beam. Electron-beam-induced temperature change is used to control the bistable states arising from the mechanical nonlinearity of the nanowire. Based on the incident electron beam position and the frequency of driven nanowire oscillation, the nanowire can be switched between its bistable states. Our results illustrate how an electron microscope provides a versatile platform not just for characterization of nanomechanical structures, but also for their active control.

**Acknowledgement.** This work was supported by the Engineering and Physical Sciences Research Council, UK (grant number EP/T02643X/1).

**Conflict of interest.** The authors state no conflict of interest.

## Supplementary Information

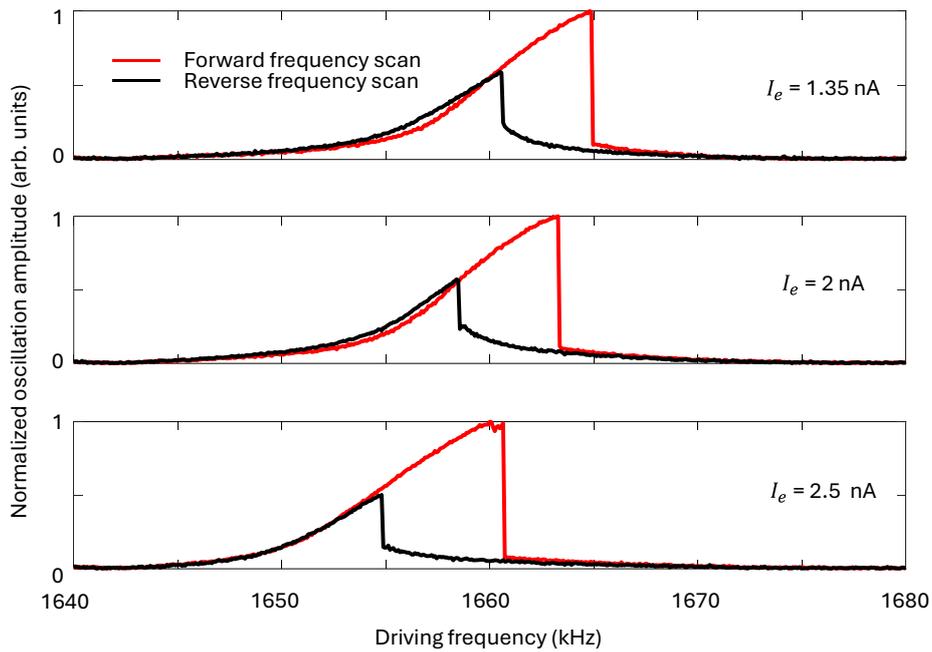

*Figure S1: Effect of the electron beam currant on the nanowire's bistable resonance. Nanowire oscillation amplitude as a function of increasing and decreasing driving frequency for a fixed driving voltage (1.5 $V_{pp}$) and different currants $I_e$ of the electron beam positioned at a distance d=100 nm from the edge of a different nanowire with width W=570 nm and otherwise the same dimensions as in the main text. The data for each currant is normalized to its own maximum.*